# Understanding Kinetic Energy paradox in Quantum Mechanics

## Yuri Kornyushin

Maître Jean Brunschvig Research Unit, Chalet Shalva,
Randogne, CH-3975

*A concept of Kinetic Energy in Quantum Mechanics is analyzed. Kinetic Energy is not zero in many cases where there are no motion and flux. This paradox can be understood, using expansion of the wave function in Fourier integral, that is on the basis of virtual plane waves.*

## 1. Introduction

Let us write down Schrödinger equation for a free particle [1]:

$$-(\hbar^2/2m)\Delta\psi = E\psi, \qquad (1)$$

where $\hbar$ is Planck constant, divided by $2\pi$, $m$ is the mass of a particle, $\Delta$ is Laplace operator, $\psi$ is a wave function, and $E$ is the energy of the stationary state of a particle. A well-known solution of Eq. (1) is a plane wave [1], $\psi(x) = \text{const}[\exp i(\mathbf{k},\mathbf{r})]$, where $i$ is the imaginary unit, and $\mathbf{k}$ is the wave vector. This solution corresponds to a non-localized free particle. The energy of this particle is well known to be $E(k) = \hbar^2 k^2/2m$ [1]. This is a Kinetic Energy of a non-localized particle [1].

## 2. Kinetic Energy in general case

Kinetic Energy is the energy of a motion [1]. A matter flux and electric current accompany the motion of a charged matter. No motion without a flux of a matter exists. In many of the bound states of any particle, e.g., electron, the wave functions are real, they contain no imaginary part [1]. Real wave function yields no flux and no electric current for a charged particle [1]. As an example let us take an electron in a ground state of a hydrogen atom. Corresponding wave function is [1] $\psi(r) = (g^3/\pi)^{1/2}\exp{-gr}$, where $g = me^2/\hbar^2$ and $r$ is the distance from proton [1].

The operator of the momentum is $-i\hbar\nabla$, wehere $\nabla$ is gradient operator. The Kinetic Energy $T$ operator is considered to be the operator of the momentum in square, divided by $2m$ [1]. This operator $-(\hbar^2/2m)\Delta$, averaged over the whole space, is regarded to be an averaged Kinetic Energy [1]. From this follows that for the electron in a hydrogen atom $T = me^4/2\hbar^2$. So we have here a very essential Kinetic Energy without any motion [2]. This paradox can be explained using virtual plane waves expansion.

The operator $-(\hbar^2/2m)\Delta$, averaged over the whole space gives some energy. This energy is not zero because the wave function is inhomogeneous; it depends on coordinates. Let us expand the wave function into Fourier integral:

$$\psi(r) = \int a_{\mathbf{k}}[\exp i(\mathbf{k},\mathbf{r})]d\mathbf{k}. \tag{2}$$

The expansion coefficients are normalized, so that

$$\int a^*_{\mathbf{k}} a_{\mathbf{k}} d\mathbf{k} = 1. \tag{3}$$

The Kinetic Energy operator $-(\hbar^2/2m)\Delta$, averaged over the whole space, is the Kinetic Energy of a system. Applying it to Eq. (2) we have:

$$T = (\hbar^2/2m) \int k^2 a^*_{\mathbf{k}} a_{\mathbf{k}} d\mathbf{k}. \tag{4}$$

Equation (4) shows that the Kinetic Energy is a sum of kinetic energies of constituting plane waves with the appropriate weight.

## 4. Discussion

The concept of *Kinetic Energy* in Non-Relativistic Quantum is Mechanics reconsidered here. Presently used concept gives non-zero values for many cases with no motion and flux. This paradox can be understood, using expansion in virtual plane waves. This expansion presents the Kinetic Energy as a sum of partial Kinetic Energies of constituting virtual plane waves.